\documentstyle[A4,11pt]{article}
\newcommand{\beqn}{\begin{eqnarray*}}
\newcommand{\eeqn}{\end{eqnarray*}}

\newcommand{\nutau}{${\rm \nu_{\tau} \ }$}

\newcommand{\taumu}{${\nu_\mu} {\leftrightarrow} {\nu_\tau} {\mbox{
oscillations}\ }$}

\newcommand{\too}{${\rm \tau \ }$}

\newcommand{\etocm}{E^{c.m.}_{\tau}}
\newcommand{\enucm}{E^{c.m.}_{\nu}}
\newcommand{\Mtoo}{ m_{\tau}}
\newcommand{\deltato}{\delta_{\tau}}
\newcommand{\deltan}{\delta_N}

\newcommand{\pnu}{p_{\nu}}
\newcommand{\pnub}{p_{\cc{\nu}}}
\newcommand{\pto}{p_{\tau}}
\newcommand{\ptob}{p_{\tau^+}}
\newcommand{\pq}{p_{q}}

\newcommand{\pqb}{p_{\cc{q}}}
\newcommand{\pqp}{p_{q'}}
\newcommand{\pqbp}{p_{\cc{q'}}}

\newcommand{\cc}[1]{\overline{#1}}
\newcommand{\weakf}{\frac{G_F}{\sqrt{2}}}
\newcommand{\lh}{(1-\gamma^5)}

\newcommand{\demi}{\frac{1}{2}}
\newcommand{\gup}[1]{\gamma^{#1}}
\newcommand{\gdo}[1]{\gamma_{#1}}

\begin{document}
\title{\Large Cross-section and polarization of neutrino-produced $\tau$'s  made simple}
\author{Jean-Michel L\'evy \thanks{Laboratoire de Physique Nucl\'eaire et de Hautes Energies, 
CNRS - IN2P3 - Universit\'es Paris VI et Paris VII, Paris.   \it Email: jmlevy@in2p3.fr}}
\pagenumbering{arabic}
\sloppy
\maketitle
\begin{abstract}
Practical formulae are derived for the cross-section and polarization vector of the
$\tau$ lepton produced in deep inelastic neutrino-nucleon scattering in 
the frame of the simple quark-parton model.
\end{abstract}
\newpage
\begin{center}
\Large{\underline{Introduction.}}
\end{center}
\par
The increasing amount of evidence for \taumu entails that some experiments being prepared 
should actually 'see' the \too produced by C.C. \nutau interaction. 
Accordingly, papers have appeared of late with the purpose
of giving the experimental groups the necessary tools to prepare their Monte-Carlo
and the analysis of their future data. One of the themes of these papers is of
course the calculation of the \too production cross-section, both differential and
integrated, and that of the \too polarization which is necessary to predict the
angular distribution of the decay products into various channels, see e.g. 
\cite{newref1}\cite{newref2}\cite{newref3} \\
However, not all these papers give simple recipes which can easily be implemented in
simulation programs. 
We therefore thought it useful to write this note which describes the way 
we calculated the cross-section and polarisation
and implemented them in the simulation which served many years ago to prepare the 
proposal for the NOMAD experiment \cite{propo}. The following is
essentially an improvement over an unpublished internal note \cite{intern} \\

We shall only treat the deep inelastic scattering (DIS) case in the quark-parton model. 
This makes things particularly simple since quarks are then described by Dirac spinors. 
However, the procedure can be straightforwardly extended to quasi-elastic or resonance 
production, through the introduction of form factors. 
These complicate the expression for the hadronic current, but the main result, viz
the extraction of the \too polarization 4-vector can be done by mimicking what follows.

Formulae for C.C. neutrino interactions without neglect of the lepton mass
were first published in \cite{albright} (see also \cite{albright3}). 
However, they were written in terms of the scaling functions
$F_i's$ and no connection was given with the basic quark/anti-quark densities
given by the available structure functions library packages.
Moreover, \cite{albright} gives a 'polarized' part of the cross-section which is 
not very useful for Monte-Carlo, whereas the polarization vector to be used in 
decay matrix elements is not explicitly written. 
It could be extracted from the given formulae but the result would only be an
average with respect to the various elementary nuclear constituents. In using
Monte-Carlo of the Lund type, where proper simulation of the hadronic final
state requires knowledge of the type of quark/anti-quark which has been hit, it
is more appropriate to derive a polarization for each case.
 
In the spirit of the impulse approximation, we shall therefore start from the
elementary interactions and derive what should represent the observable
cross-section and polarization by summing and/or averaging. 
We also make explicit the derivation of the kinematical limits in $x$ and $y$
which are only stated in the above mentionned papers and can be used as training 
for the students.
\par
Notations and basic formulae are listed in section 1; in section 2 we derive the kinematical
bounds for neutrino-nucleon C.C. scattering with a massive lepton
in the final state. Going over to the quark-parton model, the elementary sub-processes cross-sections 
and polarization 4-vectors for the lepton  are calculated in section 3. Section 4 sums up these results 
with a neutrino-nucleon cross-section formula and an average lepton polarization. 
\section{Notations}
The reaction to be described is \nutau $N \rightarrow $ \too $X$.\\
Four momenta : $k=p_{\nu}, p = p_N, p_{\tau}, q = p_{\nu}-p_{\tau} $\\
Masses : $\Mtoo$ for \too , $M$ for target nucleon, $W$ for $X$.\\
Kinematical variables:\\
C.M.S. energy squared: $s = M^2+2ME$ with $E$ the neutrino energy in the target rest
frame.\\
Four-momentum transfer squared: $Q^2=-q^2$ \\
Squared mass of final hadronic system: $W^2=(q+p)^2$ \\
Bjorken's variables $x$: $x=Q^2/2p.q \;\;$  $y=p.q/p.k$\\
C.M.S. scattering angle: $\theta$\\
Useful relations:\\
\begin{eqnarray}
&&Q^2=xy(s-M^2)=-\Mtoo^2+2\enucm(\etocm-p_{\tau}^{c.m.}cos \theta)\\
&&W^2=s+\Mtoo^2-2\etocm \sqrt{s}\\
&&x=\frac{Q^2}{W^2-M^2+Q^2}\\
&&y=\frac{W^2-M^2+Q^2}{s-M^2}=\frac{W^2-M^2}{(1-x)(s-M^2)}
\end{eqnarray}
\section{$\nu-N \rightarrow l-X$ kinematics with non zero lepton mass.}
From (3) and $W^2 \geq M^2$ one finds $x \leq 1$ as usual. Replacing 
$W^2$ taken from (2) into (4) leads to the following expression for  
$\etocm$: \begin{equation}
\etocm = \frac{\Mtoo^2+(s-M^2)(1-y+xy)}{2\sqrt{s}} \label{etoc} \end{equation}
On the other hand, the second of relations (1) and $|cos\theta | \leq 1$ yield:
$$|xy(s-M^2) + \Mtoo^2  -2\etocm\enucm| \leq
2p_{\tau}^{c.m.}\enucm$$ 
Now define $\deltato=\frac{\Mtoo^2}{s-M^2},
\; \deltan=\frac{M^2}{s-M^2}$, and use 
$\enucm=\frac{s-M^2}{2\sqrt{s}}$ to rewrite this:
$$|xy+\deltato-\frac{\etocm}{\sqrt{s}}|\leq \frac{p^{c.m.}_{\tau}}{\sqrt{s}}$$
or letting $h \stackrel{def}{=}xy+\deltato$ and squaring:$$h^2-2h\frac{\etocm}{\sqrt{s}}+
\frac{\Mtoo^2}{s} \leq 0$$
Using (\ref{etoc}) for $\etocm$ and the definitions of $\deltato,
\deltan$, this can be transformed to:
\begin{equation}
h^2\deltan -h(1-y) +\deltato \leq 0 \label{nlimh} \end{equation} or, by re-expressing $y$ as
function of $h$:
\begin{equation}
(1+x\deltan)h^2-(x+\deltato)h+x\deltato \leq 0
\label{limh}
\end{equation}
Both inequalities (\ref{nlimh}) and (\ref{limh}) lead to the kinematical limits in terms of $x$ and $y$,
albeit in slightly different forms.
Solving the first one for $h$ will give limits on $x$ as a function of $y$ and conversely for the second. 
This later inequality leads to the limits quoted in \cite{albright} as follows:  
(\ref{limh}) is possible only if: $\Delta =(x-\deltato)^2-4x^2\deltato
\deltan \geq 0$, i.e.:
\begin{equation}
|x-\deltato| \geq 2x\sqrt{\deltato \deltan}
\label{firstx}
\end{equation}
And the limits on $y$ for given $x$ are then found from  the roots of the trinomial in (\ref{limh}):
\begin{equation}
\frac{x-\deltato-2x\deltato\deltan-\sqrt{\Delta}}{2(1+x\deltan)x} \leq y \leq
 \frac{x-\deltato-2x\deltato\deltan+\sqrt{\Delta}}{2(1+x\deltan)x}
\label{limy}
\end{equation}
Clearly, one must have $x > \deltato$ so that (\ref{firstx}) is to be understood as :
\begin{equation}
x \geq x^{min} = \frac{\deltato}{1-2\sqrt{\deltato \deltan}}
=\frac{\deltato}{1-2\frac{\Mtoo M}{s-M^2}}=\frac{\deltato}{1-\frac{\Mtoo}{E}}
\label{limx}
\end{equation}
Note that the upper limit for $x$ derived from (\ref{nlimh}) is irrelevant (above 1) but that as soon
as one demands a minimum value for $W$ above $M$ (DIS should mean $W > M_{\Delta}$ at least), then relation 
(4) entails an upper limit on $x$ as a function of $y$

\section{Elementary interactions.}
To derive formulae for the differential cross-section and polarization, we use
the simplest quark-parton model and then sum over partonic contributions. 
\subsection{Matrix elements.}
The kinematics is $\pnu + \pq = \pto + \pqp$ \footnote{\bf we will use this notation
even for anti-quarks and anti-neutrinos except in the final x-section formulae}
and use shall be made of the $\tau$ polarization four-vector $S$ \\
The reaction amplitude is:
$$\weakf \bar{u}(\tau)\gup{\alpha} \lh u(\nu) \bar{u}(q')\gdo{\alpha}\lh u(q)
B.W.$$ where $$B.W.= \frac{M^2_W}{M^2_W-q^2-i\Gamma_W M_W}$$ is a W's
propagator correction to the pure Fermi amplitude, where terms arising from the longitudinal 
($\frac{q^{\alpha}q^{\beta}}{M^2_W} $) part have been neglected. \\
Squaring and introducing the density matrices one gets:
$$\frac{G^2_F}{2} Tr(\rho_{\tau} \gup{\alpha} \lh \rho_{\nu}\gup{\beta} \lh )
Tr(\rho_{q'} \gdo{\alpha} \lh \rho_q \gdo{\beta} \lh ) $$
Here $\rho_{\tau}=\demi (\gamma \cdot p_{\tau}+ \Mtoo)(1+\gup{5} \gamma \cdot
S)$ but the other density matrices are summed (averaged) over polarizations and  
$|B.W.|^2$ is understood here and in what follows.\\
The first trace is the so-called leptonic tensor, which, with $\rho_{\nu} = \gamma \cdot k$ ($\pnu = k$ 
here, for ease of notation)
reads explicitly: $$L^{\alpha,\beta}_{\nu} = 4(L^{\alpha}k^{\beta}+L^{\beta}k^{\alpha}-g^{\alpha \beta}L\cdot 
k -i\epsilon^{\mu \alpha \nu \beta}L_{\mu}k_{\nu})$$ where $L\stackrel{def}{=}k-m_{\tau}S$ \\
For anti-neutrinos, $L$ should read $k+m_{\tau}S$ and the sign of the antisymmetric part should be reversed.

Taking the trace of the quark tensor and contracting,  one finds:
\begin{equation}
|T|^2 = 32 G_F^2 \pnu \cdot \pq (\pto - \Mtoo S) \cdot \pqp
\label{nuq}
\end{equation}
We shall use the $S$ dependence of this result later. For the moment, we sum
it over polarizations to get the transition probability: 

$$64 G_F^2 \pnu \cdot \pq \pto  \cdot \pqp$$ 
\par
For an anti-quark target, one gets instead of (\ref{nuq}):
\begin{equation}
32 G_F^2 \pnu \cdot \pqbp (\pto - \Mtoo S) \cdot \pqb 
\label{nuqb}
\end{equation}
so that the roles of the initial and final partons are permuted.
Finally for an anti-neutrino going over into $\tau^+$ one finds:
\begin{equation}
32 G_F^2 \pnub \cdot \pqp (\pto + \Mtoo S) \cdot \pq 
\label{nubq}
\end{equation}
and
\begin{equation}
32 G_F^2 \pnub \cdot \pqb (\pto + \Mtoo S) \cdot \pqbp
\label{nubqq}
\end{equation}
for a quark and anti-quark target respectively.
\subsection{Cross-sections.}
\label{xsec}
The differential cross-section for each of the elementary processes considered
above is:$$
d\sigma = \frac{1}{F}|T|^2 (2
\pi)^{-2}\delta^4(\pqp+\pto-\pq-\pnu)\frac{d^3\pqp}{2\pqp^0}
\frac{d^3\pto}{2\pto^0}$$ 
with $F$ the M\"oller flux factor and $|T|$ is the matrix
element computed above. Standard manipulations and integration 
with respect to the $\tau$ azimuthal angle reduce this expression to $$d\sigma = \frac{1}
{8\pi F}|T|^2 dy$$ with $y$ defined in section $1$ as the leptonic fractional energy
loss in the target \underline{nucleon} rest-frame. The flux factor is simply
$F=4 \pnu \cdot \pq$. 
\subsection{Polarization \label{Polar}}
Here shall be found the main simplification of our presentation w.r.t. others.
Setting up a polarization basis is useful for testing symmetries or conservation 
laws in the production process, but is of no use for simulations 
which only require the $\tau$ polarization four-vector in any specified frame, to
be fed, for example, in a decay matrix element.\\
The calculation is most simply done by following the reasoning of (\cite{L.L.}): 
the squared amplitudes written above are proportional to the probabilities for
finding the final $\tau^{\pm}$ in a given state of polarization described by the
four-vector $S$. If the true polarization from the production process is $S_f$,
then those probabilities are equally found by projecting the true density
matrix $1+\gup{5}\gamma \cdot S_f$ on $1+\gup{5}\gamma \cdot S$ which represents
the given state for which we want the probability. \\
In other words, $|T|^2 \propto Tr(1+\gup{5}\gamma \cdot S)(
1+\gup{5}\gamma \cdot S_f) \propto 1-S \cdot S_f$ \\
Hence we find for the $\nu - q$ case:
$$S \cdot S_f = \frac{S \cdot \pqp \Mtoo}{\pto \cdot \pqp}$$
Since $S_f \cdot \pto =0$ we cannot simply invoke the arbitrariness of $S$ to
cross it away but we must allow for a term proportional to $\pto$:
$$ S_f =  \frac{\Mtoo \pqp}{\pto \cdot \pqp} + \lambda \pto$$
$\lambda$ is now determined by projecting this equality onto $\pto$ and found
to be $-\frac{1}{\Mtoo}$ so that the final result reads:
\begin{equation}
S_f =  \frac{\Mtoo \pqp}{\pto \cdot \pqp} - \frac{\pto}{\Mtoo} \label{pol}
\end{equation}
In the $\tau$ rest-frame we see that the time component $S_f^0 = 0$ as it
should be and that the space part (the polarization in its usual sense) is: $$
\vec{{\cal P}_f} = \frac{\vec{\pqp}}{E_{q'}} $$ 
For the three other cases one finds:\\
\begin{displaymath}
\begin{array}{|c | c | c|}\hline
\nu-\bar{q} & \bar{\nu}-q & \bar{\nu}-\bar{q} \\ \hline
\vec{{\cal P}_f} = \frac{\vec{\pqb}}{E_{\cc{q}}}&
\vec{{\cal P}_f} = -\frac{\vec{\pq}}{E_{q}}&
\vec{{\cal P}_f} = -\frac{\vec{\pqbp}}{E_{\cc{q'}}}\\ \hline
\end{array}
\end{displaymath}\\

It is important to note that what has been done here with the simple quark tensor can be adapted
to more complicated forms of the hadronic current and of the tensor built from it. For example, if
$\tau$ is produced through a quasi-elastic C.C. interaction, the matrix element of the hadronic current
between the neutron and the proton states is described by six form factors (two of which are zero in
this case) but exactly the same procedure can be used to identify $S_f$ in terms of 4-momenta, kinematical
invariants and these form factors.   
\section{Interactions with Nucleons}
\subsection{\nutau - Nucleon cross-section}
The elementary cross-sections written in \ref{xsec} read explicitly
$$d\sigma(\nu -q) = \frac{2G_F^2}{\pi\pnu\cdot\pq}(\pnu\cdot\pq \  \pto\cdot\pqp) dy$$
$$d\sigma(\nu -\cc{q}) = \frac{2G_F^2}{\pi\pnu\cdot\pqb}(\pnu\cdot\pqbp \  \pto\cdot\pqb) dy$$
$$d\sigma(\cc{\nu} -q) = \frac{2G_F^2}{\pi\pnub\cdot\pq}(\pnub\cdot\pqp \  \ptob\cdot\pq) dy$$
$$d\sigma(\cc{\nu} -\cc{q}) = \frac{2G_F^2}{\pi\pnub\cdot\pqb}(\pnub\cdot\pqb \  \ptob\cdot\pqbp) dy$$
Let $\xi$ be the nucleon momentum fraction carried by the struck (anti-) quark:
$\pq = \xi p$. Energy-momentum conservation says that:
\begin{eqnarray}
&\pnu + \xi p = \pto + \pqp & \label{eq:e-m-c}\\ 
{\rm or \ }\ & q+\xi p = \pqp & \label{eq:qemc}
\end{eqnarray}
by  squaring (\ref{eq:qemc}) we get:
\begin{eqnarray}
& 2(\xi-x)p \cdot q + \xi^2 M^2 = m_{q'}^2 & \label{eq:xxi}
\end{eqnarray}
In Bjorken's limit, $q^2 \rightarrow \infty, p \cdot q \rightarrow \infty$
(\ref{eq:xxi}) shows that $\xi = x$ \\
Squaring now (\ref{eq:e-m-c}), combining with (\ref{eq:xxi}) and using $\xi = x$ yields:
$$2\pto \cdot \pqp = x(s-M^2) - \Mtoo^2$$
Therefore :
\begin{equation}
\label{eq:dsq}
d\sigma(\nu -q) = d\sigma(\cc{\nu}
-\cc{q})=\frac{2G_F^2}{\pi}MEx(1-\frac{\deltato}{x})dy 
\end{equation}
Taking the scalar product of (\ref{eq:e-m-c}) by $\pnu$, using again $\xi=x$ and the 
definition of $q$ yields: $$2 \pqp \cdot \pnu =
x(s-M^2)(1-\frac{\deltato}{x}-y)$$ 
On the other hand: $$2\pq \cdot \pto = x(1-y)(s-M^2)$$ and $$F = 4 \pnu \cdot
\pq = 2x(s-M^2)$$ \\ 
Hence:
\begin{equation}
\label{eq:dsqb}
d\sigma(\nu -\cc{q}) = d\sigma(\cc{\nu}-q)=\frac{2G_F^2}{\pi}
MEx(1-\frac{\deltato}{x}-y)(1-y)dy
\end{equation}
Formulae (\ref{eq:dsq}) and (\ref{eq:dsqb}) solve the cross-section question. Multiplying
them by the appropriate quark/anti-quark distribution functions, summing over
flavors and re-introducing the $W$ boson propagator factor, one finds for a
neutrino beam:
\begin{equation}
\frac{d \sigma}{dx \ dy}\ =\ \frac{2 G^2_F E M x}{\pi}
\left[(1-\frac{\deltato}{x}) {\cal Q}(x,Q^2) + (1-\frac{\deltato}{x}-y) (1-y)
\overline{\cal Q}(x,Q^2)\right]|B.W.(Q^2)|^2 
\label{xsection}
\end{equation}
where ${\cal Q}$ and $\cc{\cal Q}$ are the appropriate mixtures of (Q.C.D.-
evolved) quark distribution functions for the nucleon which is hit.\\
For an anti-neutrino beam, the coefficients of the quark and anti-quark 
distribution functions must evidently be exchanged and the distributions
themselves adequately modified for the given nucleon, in order to take into 
account charge conservation at the constituent level.\\

Expressions (\ref{eq:dsq})(\ref{eq:dsqb}) and/or (\ref{xsection}) are directly 
usable with standard quark distribution
function libraries; from that point of vue, they are more practical than expressions
found in the quoted articles where nucleons are described by structure functions
or their scaling limits. In this same limit, the connection between the two descriptions
is easily made along the lines of what can be found, for example, in (\cite{Chengli}) 
for the electromagnetic interactions: one merely has to identify the general hadronic tensor 
with the sum of quark and antiquark tensors decomposed on the same basis; the various relations 
$F_2 = 2xF_1$, $F_4 =0$, $F_5 = F_1$ follow and after simplifications, 
formula (\ref{xsection}) is retrieved. 
\subsection{\too polarization in $\nu$-nucleon scattering \label{toPolar}}
The two terms in (\ref{xsection}) are to be interpreted as the relative
probabilities for scattering from a quark and an anti-quark in the target
nucleon. The \too polarization vector is therefore the weighted average of the
two relevant ${\cal P}_f$ of section 3.3.\\

Let us call $P_Q = (1-\frac{\delta_{\tau}}{x}){\cal Q} $ and 
$P_{\bar{Q}} = (1-y)(1-y-\frac{\delta_{\tau}}{x})\cc{\cal Q} $\\
The coefficients of $P_Q$ and $P_{\bar{Q}}$ were found in section \ref{Polar} 
to be: $\frac{\vec{\pqp}}{E_{q'}}$ and $\frac{\vec{\pqb}}{E_{\cc{q}}}$.\\
Rewriting the denominators in Lorenz-invariant form as:\\
$E_{q'} = \frac{1}{\Mtoo} \pto \cdot (q+xp) = \frac{s-M^2}{2\Mtoo} x (1-\frac{\deltato}{x})$ and 
$E_{\cc{q}} = \frac{\pto \cdot \pqb}{\Mtoo} = \frac{s-M^2}{2\Mtoo}x(1-y)$ 

we find the average $\tau$ polarization in neutrino scattering:
 $$\vec{\cal P}_{\tau} = \frac{2\Mtoo}{s-M^2}\left [(\vec{p}+\frac{\vec{q}}{x})
{\cal Q}(x,Q^2)+\vec{p}(1-\frac{\deltato}{x}-y)\cc{\cal Q}
(x,Q^2) \right ] (P_Q + P_{\bar{Q}})^{-1}$$
and in anti-neutrino scattering:
 $$\vec{\cal P}_{\tau^+} = -\frac{2\Mtoo}{s-M^2}\left
[(\vec{p}+\frac{\vec{q}}{x})\cc{\cal
Q}(x,Q^2)+\vec{p}(1-\frac{\deltato}{x}-y){\cal Q}(x,Q^2) \right ]  (P_Q + P_{\bar{Q}})^{-1}$$
with the same proviso as in the preceding subsection as to the contents of $\cal Q$ and $\cc{\cal Q}$ 

In these expressions, all 3-vectors are expressed in the \too rest-frame. Should one need $S_f$ in another
frame, one could either boost the four vector $(0, \vec{\cal P})$ or perform the same averaging directly
with the $S_f$ vectors extracted from the transition probabilities as done in section (\ref{Polar}).
It is clear that the relative probabilities are made of Lorentz invariants only and therefore valid in
any frame, as it must be physically.\\
Let us remark, however, that for a Lund-type Monte-Carlo where the struck quark is identified, no
averaging takes place at the event level and one uses directly the results of (\ref{Polar}). 

\section{Conclusion}
We have given lowest order but simple formulae for cross-section and polarization of \too 's produced
by \nutau 's charged current interactions valid in the DIS regime. We leave it to the interested reader 
to use her/his favorite quark distribution functions package to calculate and study cross-sections and 
polarizations with the help of these formulae through Monte-Carlo or numerical integration.

\newpage
\end{document}